\documentclass[nosuperscriptaddress, showpacs, twocolumn,titlepage, showkeys, nofootinbib]{revtex4-2}
\usepackage{times}
\usepackage{amsmath}
\usepackage{amssymb}
\usepackage{graphicx}
\usepackage{subfigure}
\usepackage{booktabs}
\usepackage{rotating}
\usepackage{xcolor}
\usepackage{braket,mleftright}
\usepackage{longtable}
\usepackage{bm}
\usepackage{float}

\makeatletter

\newcommand{\Rmnum}[1]{\expandafter\@slowromancap\romannumeral #1@}
\makeatother

\begin{document}

\title{Quantum version of the k-NN classifier based on a quantum sorting algorithm}

\author{L. F. Quezada}
\email{E-mail: lfqm1987@ciencias.unam.mx}
\affiliation{Huzhou University, Huzhou 313000, P.R. China}
\affiliation{Laboratorio de C\'{o}mputo Inteligente, Centro de Investigaci\'{o}n en Computaci\'{o}n, Instituto Polit\'{e}cnico Nacional, UPALM, 07700, Ciudad de M\'exico, M\'exico.}

\author{Shi-Hai Dong}
\email{E-mail: dongsh2@yahoo.com}
\affiliation{Laboratorio de Informaci\'{o}n Cu\'{a}ntica, Centro de Innovaci\'on y Desarrollo Tecnol\'ogico en C\'omputo, Instituto Polit\'ecnico Nacional, UPALM, 07700, Ciudad de M\'exico, M\'exico.}
\affiliation{Huzhou University, Huzhou 313000, P.R. China}

\author{Guo-Hua Sun}
\email{E-mail: sunghdb@yahoo.com}
\affiliation{Laboratorio de C\'{o}mputo Inteligente, Centro de Investigaci\'{o}n en Computaci\'{o}n, Instituto Polit\'{e}cnico Nacional, UPALM, 07700, Ciudad de M\'exico, M\'exico.}

\begin{abstract}
In this work we introduce a quantum sorting algorithm with adaptable requirements of memory and circuit depth, and then use it to develop a new quantum version of the classical machine learning algorithm known as k-nearest neighbors (k-NN). Both the efficiency and performance of this new quantum version of the k-NN algorithm are compared to those of the classical k-NN and another quantum version proposed by Schuld et al. \cite{Int13}. Results show that the efficiency of both quantum algorithms is similar to each other and superior to that of the classical algorithm. On the other hand, the performance of our proposed quantum k-NN algorithm is superior to the one proposed by Schuld et al. and similar to that of the classical k-NN.
\end{abstract}

\maketitle

\section{Introduction}

Machine learning is a branch of artificial intelligence which studies the methodology behind an algorithm being able to make decisions or predictions without being explicitly programmed to do so, but rather carried out automatically. This can be done based on previous related information (supervised learning), by finding structures within the available data (unsupervised learning) or by penalizing mistakes in a trial and error process (reinforcement learning).

To date, various machine learning models have been proposed. In 1951, Fix and Hodges developed the k-nearest neighbors (k-NN) algorithm \cite{Int1}, which classify a data pattern according to the most common label among its k nearest neighbors. In 1958, Frank Rosenblatt proposed the perceptron \cite{Int2}, which is the fundamental block of the multilayer perceptron, one of the most widely used class of artificial neural networks. Among the support-vector models, the support-vector clustering algorithm developed in 2001 by Ben-Hur, Horn, Siegelmann and Vapnik \cite{Int3}, is now widely used in the industry for unsupervised learning. Probability-based algorithms such as the Naive Bayes classifier \cite{Int4}, due to its simplicity and performance, has gained a place in the set of ``algorithms to beat'' in various benchmarking processes.

On the other hand, quantum computing emerged in 1980 with the work of Paul Benioff \cite{Int5}, in which a quantum model of the Turing machine was proposed. In essence, quantum computing is a new way of computing that takes advantage of collective properties of quantum systems, such as superposition and entanglement, to perform concrete calculations faster than a regular or classical computer. In 1982, Richard Feynman and Paul Benioff showed that a classical computer is not capable of simulating a quantum system without a disproportionate amount of resources \cite{Int6}, opening the door for multiple potential applications in chemistry and material science \cite{Int7,Int8,Int9,Int10}. To date, several examples of the superiority of quantum computing over classical computing have been given \cite{Int10-5}, being Shor's factorization algorithm \cite{Int11} and Grover's search algorithm \cite{Int12} two of the most famous examples. Furthermore, algorithms to achieve quantum advantage in Noisy Intermediate-Scale Quantum (NISQ) computers have also been proposed \cite{Int12-5}.

In the last couple of decades, due to the increasing applications of machine learning, and with a well-established library of quantum algorithms, the interest of part of the related scientific community has move toward the use of quantum computing for the optimization of classical machine learning algorithms, as well as the development of fully quantum algorithms with the same capabilities. In 2009, Harrow, Hassidim and Lloyd (HHL) developed a quantum algorithm for solving linear systems of equations \cite{Int12-6}, which has been widely used for new proposals in quantum machine learning, such as quantum vector support \cite{Int12-7}, quantum Hopfield neural networks \cite{Int12-8}, quantum linear regression \cite{Int12-9} and quantum data fitting \cite{Int12-10}. In 2014, Maria Schuld et al. proposed a quantum version of the k-NN algorithm \cite{Int13}; one year after, the same authors proposed a way of simulating a perceptron using a quantum computer \cite{Int14}. In 2017, Biamonte et al. show the complexity and main characteristics of various quantum machine learning algorithms \cite{Int15}. More recently, in 2021, proposals on quantum machine learning included: encoding patterns for quantum algorithms \cite{Int15-1}, financial applications \cite{Int15-2}, federated quantum machine learning \cite{Int15-3}, an algorithm for knowledge graphs \cite{Int15-4} and algorithms for drug discovery applications \cite{Int15-5}.

Hybrid proposals using both classical and quantum computing have also been made. In 2020, LaBorde et al. used a classical k-NN algorithm to find broken quantum gates in a circuit \cite{Int16}. That same year, Henderson et al. proposed the use of convolutional neural networks with a layer of random quantum circuits, essentially quantum-transforming data before processing it with a classical algorithm \cite{Int17}. In 2021, Cerezo et al. present an overview on variational quantum algorithms \cite{Int18}, Endo et al. review the basic results for hybrid algorithms \cite{Int19}, and Wurtz and Love propose integrating classical precomputation subroutines to improve the performance of variational quantum algorithms \cite{Int20}.

In this work we introduce a quantum sorting algorithm which we then use to develop a new quantum version of the k-NN algorithm. We compare the efficiency and performance of our proposal with both the classical k-NN algorithm and the quantum version developed by Schuld et al. \cite{Int13}. The paper is organized as follows: in sections \ref{C_kNN} and \ref{S_kNN} we respectively give a brief description of the classical k-NN algorithm and Schuld et al. quantum version; in section \ref{MP} we introduce a quantum sorting algorithm, which we then use to develop our proposed quantum version of the k-NN algorithm in section \ref{MP_kNN}; results between the comparison of the three algorithms are presented in section \ref{ResD}, in which we also discuss them; lastly, in section \ref{C} we give some conclusions.

\section{Classical k-NN algorithm} \label{C_kNN}

The k-nearest neighbors algorithm (k-NN) is a supervised machine learning algorithm for pattern classification. Given a pattern $x^{\omega}$ for classification and a training set $E$, the k-NN algorithm works by calculating the distance between $x^{\omega}$ and each of the elements in $E$, and assigning the majority class among the nearest $k$ patterns. Due to its simplicity and good performance, the k-NN algorithm is widely used with all sorts of datasets \cite{Ck-NN1,Ck-NN2,Ck-NN3}, modified versions of it have been presented \cite{Ck-NN4,Ck-NN5}, and even an efficient approach for IoT devices has been proposed \cite{Ck-NN6}.

Like many machine learning algorithms for classification, the operation of k-NN can be divided into two stages: the training stage and the classification stage. Consider the training set of $n$-dimensional patterns $E = \left \lbrace (x^{1}, c^{1}), \dots, (x^{N}, c^{N}) \right \rbrace $, where $x^{i}$ denotes the i-th pattern and $c^{i}$ its corresponding class. k-NN belongs to a type known as lazy algorithms, which means that its training stage consists solely in storing $E$ and having it available. On the other hand, the classification stage involves calculating the distances between the pattern to be classified $x^{\omega}$ and all the patterns in $E$. This requires to define a metric, that is, a distance function $d$ that satisfies the following conditions for any $x,y,z$:
\begin{enumerate}
	\item $d(x,y) \in \left[ 0,\infty \right)$
	\item $d(x,y) = 0 \iff x = y$
	\item $d(x,y) = d(y,x)$
	\item $d(x,z) \leq d(x,y) + d(y,z)$
\end{enumerate}

With a distance function $d$ already defined, the distances between $x^{\omega}$ and all the patterns in $E$ are calculated, forming the array $D = \left[ d \left( x^{\omega}, x^{1} \right), \dots, d \left( x^{\omega}, x^{N} \right) \right]$. Then, the $k$ least values in $D$ are taken, which correspond to the $k$ nearest patterns to $x^{\omega}$. Lastly, $x^{\omega}$ is assigned the majority class among those $k$ patterns.

It should be mentioned that, when the training set has only two classes, it is enough to use an odd number $k$ to avoid ties; however, when there are more than two classes, the possibility of a tie occurring exists for any value of $k$. Thus, it is necessary to define a rule to break these ties. A commonly used rule is to use the least number among the tied ones.

Regarding the complexity of the algorithm, if the distances between $x^{\omega}$ and the patterns in the training set are calculated, stored, and then used to find the least $k$, the complexity is $O(nN + kN)$. On the other hand, if all distances are calculated $k$ times, each time searching for a new minimum not previously found, then the complexity is $O(knN)$.

\section{Quantum version of the k-NN algorithm proposed by Schuld et al. \cite{Int13}} \label{S_kNN}

	Consider a pattern $x^{\omega}$ to be classified and a training set of binary $n$-dimensional patterns $E = \left \lbrace (x^{1}, c^{1}), \dots, (x^{N}, c^{N}) \right \rbrace $. Then, the quantum k-NN algorithm proposed by Schuld et al. \cite{Int13} transforms the initial state
	\begin{equation}
	\ket{\psi_{i}} = \frac{1}{\sqrt{N}} \sum_{i=1}^{N} \Ket{x^{\omega},x^{i},c^{i},0},
	\end{equation}
	into the final state
	\begin{align}
	\notag \ket{\psi_{f}} =&
	\, \frac{1}{\sqrt{N}} \sum_{i=1}^{N} \cos\left(\frac{\pi}{2n}d_{h}\left( x^{\omega}, x^{i} \right) \right) \Ket{x^{\omega},d^{\, i},c^{i},0}
	\\
	&+ \frac{1}{\sqrt{N}} \sum_{i=1}^{N} \sin\left(\frac{\pi}{2n}d_{h}\left( x^{\omega}, x^{i} \right) \right) \Ket{x^{\omega},d^{\, i},c^{i},1},
	\end{align}
	where $d_{h}\left( x^{\omega}, x^{i} \right)$ denotes de Hamming between $x^{\omega}$ and $x^{i}$, and
	\begin{equation}
	\label{distance} d^{\, i}_{j} =
	\begin{cases}
	1 \quad \text{if } x^{\omega}_{j} = x^{i}_{j} \\
	0 \quad \text{if } x^{\omega}_{j} \neq x^{i}_{j}
	\end{cases}
	\end{equation}
	Notice that, due to the cosine in the first sum, the probability of finding the last qubit in the state $\ket{0}$, is high for patterns near $x^{\omega}$; while, due to the sine in the second sum, the probability of finding the last qubit in the state $\ket{1}$, is high for patterns far from $x^{\omega}$. Given that the relevant patterns are the ones near $x^{\omega}$, the last qubit is sought to be measured in the state $\ket{0}$; the probability of doing so is given by
	\begin{equation}
		P_{0} = \frac{1}{N} \sum_{i=1}^{N} \cos^{2} \left(\frac{\pi}{2n}d_{h}\left( x^{\omega}, x^{i} \right) \right).
	\end{equation}

	Schuld et al. define a threshold $T>k$ and propose to run the algorithm $T$ times, each time measuring the last qubit:
	\begin{itemize}
		\item If the result of the measurement is $0$, then the qubit (or qubits) containing the class information is also measured. The class measurement's result is stored and considered as one of the $k$ needed.
		\item If the result of the measurement is $1$, then the run is discarded but counted as one of the $T$ runs to be performed.
	\end{itemize}
	In the first scenario; that is, if the last qubit is found in the state $\ket{0}$, the probability of obtaining an arbitrary class $c$ is given by
	\begin{equation}
	\label{prob_class_s} P(c) = \frac{1}{P_{0} N} \sum_{i \, | \, x^{i}\in c} \cos^{2} \left(\frac{\pi}{2n}d_{h}\left( x^{\omega}, x^{i} \right) \right).
	\end{equation}

	The whole process ends in one of the following two possible ways:
	\begin{itemize}
		\item The threshold $T$ is exceeded without having obtained $k$ class' candidates. In this case, the pattern $x^{\omega} $ is considered as unclassifiable, or as an error in a validation process.
		
		\item $k$ class' candidates are obtained with less or exactly $T$ runs of the algorithm. In this case, $x^{\omega}$ is assigned the majority class among those $k$ class' candidates.
	\end{itemize}
	
	It is worth mentioning that, as in the classical algorithm, it is necessary to have a rule for when there exist ties in the selection of the majority class among the $k$ candidates. However, unlike the classical algorithm, the value of $k$ is not necessarily restricted to the number of patterns in the training set, as this value can now be interpreted as the number of samples to be obtained in a random experiment with replacement, where the probabilities of measuring each class are given by the expression \eqref{prob_class_s}.
	
	Regarding the complexity of the algorithm, this in principle can be considered to be $O(nT)$. Nevertheless, an arbitrary initialization of a quantum circuit in an $N$-element superposition of $n$ qubits is $O(nN)$ \cite{Sk-NN1}; thus, if we consider the initialization, the complexity increases to $O(nTN)$. Ways for getting around this initialization problem are actively being investigated. Schuld et al., for example, mentioned the possibility of receiving the initial state from a quantum memory. Other proposals have also been given, such as efficient initialization protocols \cite{Sk-NN2}.

\section{The (m,p) Sorting Algorithm} \label{MP}

In this section we propose a quantum sorting algorithm. The specific task of this algorithm is to sort the elements of an array via the amplitude of its elements in a quantum superposition. It is worth mentioning that this task is different from the one performed by classical sorting algorithms and some quantum versions \cite{MP1,MP2,MP3}, in which the aim is to retrieve the elements of an array in order; the purpose of the quantum algorithm proposed in this section is not to retrieve the elements in a superposition, but rather to order them via their amplitude, that is, match the amplitude of each element to its order in the array.

Consider the array whose elements we would like to sort in a superposition to be $Y = \left[ y^{1}, y^{2}, \dots, y^{N} \right]$ with $y^{i} \in \left\lbrace 0, 1 \right \rbrace^{n}$ for all $i$, and suppose $\left \lbrace 0, 1 \right \rbrace^{n}$ has an strict total order relation; that is, a relation ``$<$'' such that for any $x, y, z \in \left \lbrace 0, 1 \right \rbrace^{n}$, the following conditions are satisfied:
\begin{itemize}
	\item $\left( x < y \right)$ $\vee$ $\left( y < x \right)$ (total)
	
	\item $\neg \left( x < x \right)$ (irreflexive),
	
	\item $ x < y $ $\Longrightarrow$ $\neg \left( y < x \right)$ (asymmetric),
	
	\item $\left( x < y \right)$ $\wedge$ $\left( y < z \right)$ $\Longrightarrow$ $x < z$ (transitive).
\end{itemize}
The characteristic function of this relation can then be defined as:
\begin{align}
	\notag	&f:\left \lbrace 0, 1 \right \rbrace^{2n} \longrightarrow \left\lbrace 0,1 \right\rbrace
	\\
	&f\left(x, z \right) = 
	\begin{cases}
		1 \qquad \text{if } x < z,\\
		0 \qquad \text{if } \neg \left( x < z \right).
	\end{cases}
\end{align}
In general, given $m \in \left\lbrace 2, \dots , N \right\rbrace$, this definition can be extended to
\begin{align}
	\notag \label{gcf}	&f_{m}:\left\lbrace 0, 1 \right\rbrace^{nm} \longrightarrow \left\lbrace 0,1 \right\rbrace
	\\
	\notag &f_{m}\left(z^{1}, \dots ,z^{m} \right) =  \delta^{1 \, \cdots \, m} \\
	\text{where } &\delta^{1 \, \cdots \, m} = \begin{cases}
		1 \qquad \text{if } z^{1} < z^{2} < \cdots < z^{m},\\
		0 \qquad \text{other cases.}
	\end{cases}
\end{align}
Notice that the elements in $Y$ are subject to the order relation in $\left\lbrace 0, 1 \right \rbrace^{n}$, and that the generalized characteristic function in equation \eqref{gcf} allows us to discern if any length-$m$ array of elements in $Y$ is ordered or not. Hence, the generalized characteristic function serves two main purposes: binarize the set relation to be able to encode it in a qubit, and construct a unitary oracle $U_{f_{m}}$ that allows us to query the order of any given length-$m$ array of elements in $Y$.

The $(m,p)$ algorithm is then as follows:

\begin{figure}[t]
	\begin{centering}
		\includegraphics[width=0.99\linewidth]{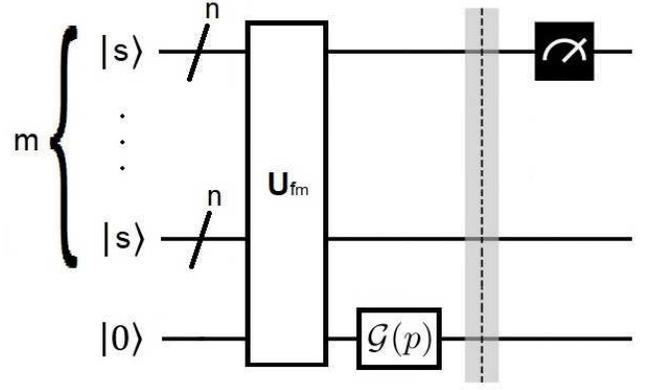}
		\par\end{centering}
	\caption{\label{fig:1} Quantum circuit of the $(m,p)$ sorting algorithm.}
\end{figure}

\begin{figure*}[t]
	\begin{centering}
		\includegraphics[width=0.47\linewidth]{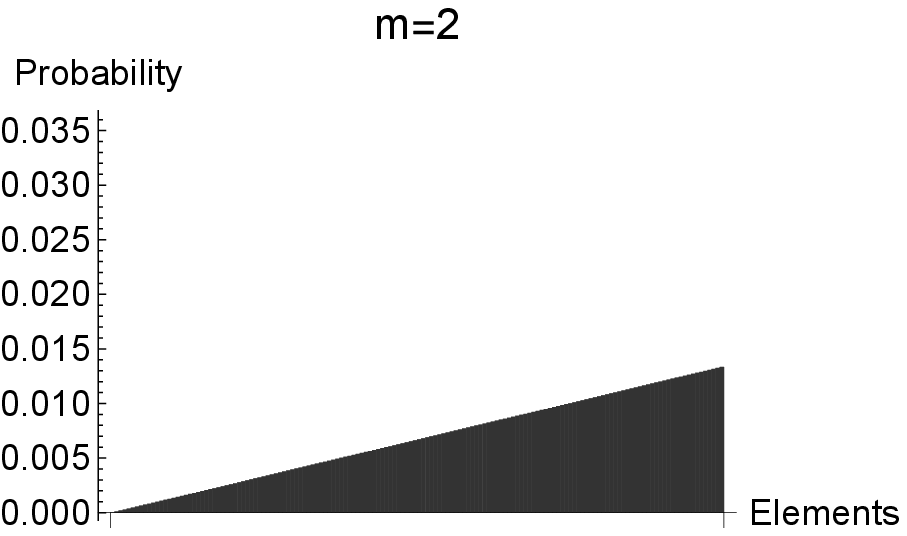}
		\quad
		\includegraphics[width=0.47\linewidth]{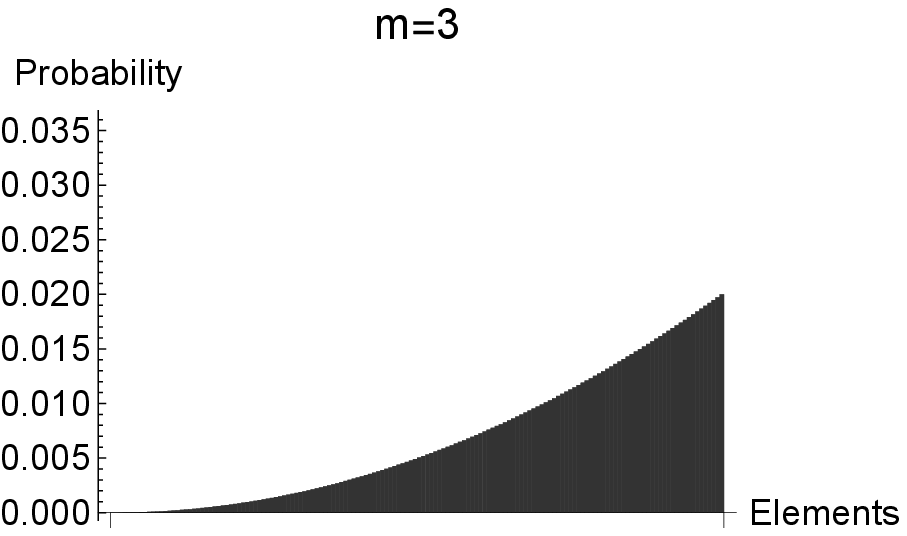}
		
		\vspace{1cm}
		
		\includegraphics[width=0.47\linewidth]{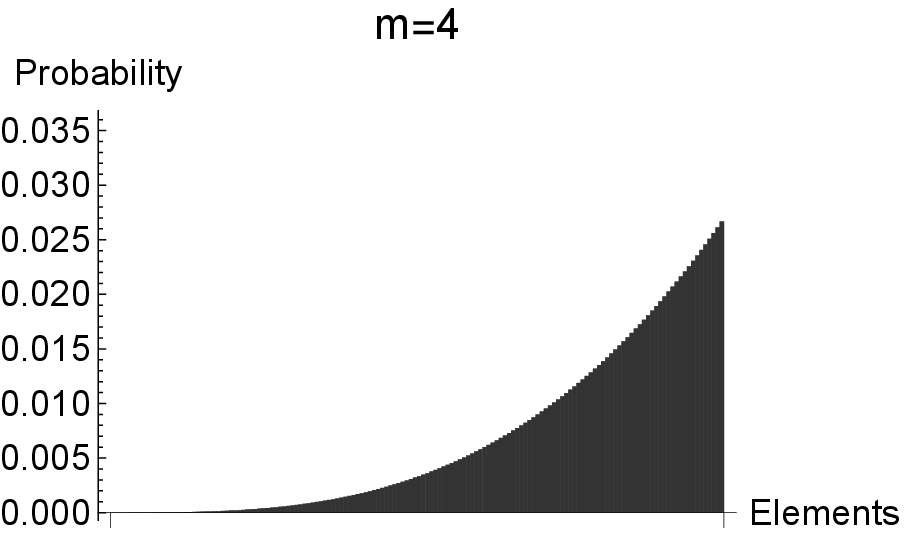}
		\quad
		\includegraphics[width=0.47\linewidth]{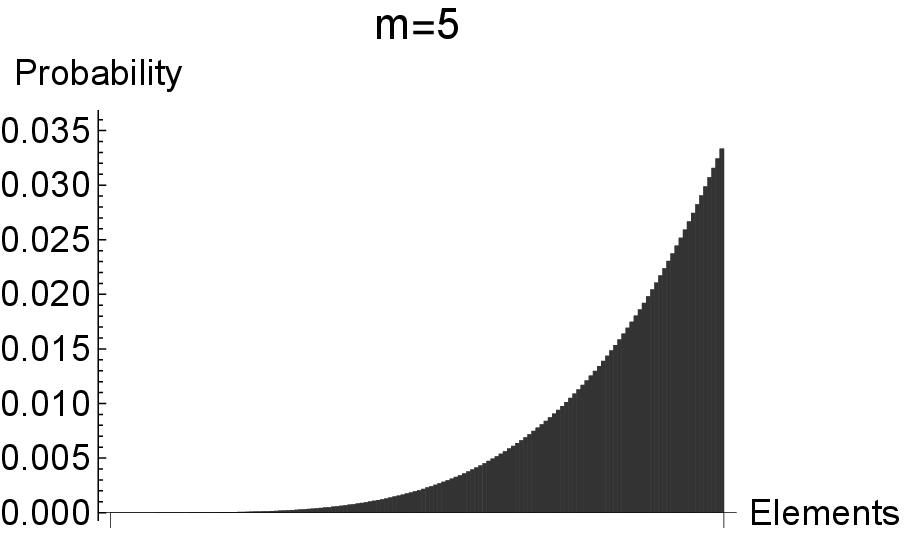}
		\par\end{centering}
	\caption{\label{fig:2} Probability distribution of the elements in $Y$ present in the state $\Ket{\phi_{1}}$ for $m=2,3,4,5$ and $N=150$. Elements in each chart are arranged from largest (left) to smallest (right).}
\end{figure*}

\begin{enumerate}
	\item Prepare the initial state 
	\begin{equation}
		\ket{\psi_{1}} = \ket{s}^{\otimes m} \otimes \ket{0},
	\end{equation}
	where $\ket{s} = \displaystyle\frac{1}{\sqrt{N}} \sum_{i=1}^{N} \Ket{y^{i}}$.
	
	\item Apply the unitary operator $U_{f_{m}}$; that is, the one that implements $f_{m}$:
	\begin{align}
		\notag	\ket{\psi_{2}} &= U_{f_{m}} \ket{\psi_{1}}
		\\
		&= \frac{1}{\sqrt{N^{m}}} \sum_{i_{1} \cdots i_{m}} \Ket{y^{i_1},\dots,y^{i_m}, \delta^{i_{1} \cdots i_{m}}}.
	\end{align}
	The state $\ket{\psi_{2}}$ can also be written as
	\begin{align}
		\ket{\psi_{2}} = \cos\theta  \ket{\phi_0} + \sin\theta \ket{\phi_1},
	\end{align}
	where
	\begin{align}
		\ket{\phi_0} = \frac{1}{\sqrt{\nu}} \sum_{i_{1} \cdots i_{m}} \Ket{y^{i_1},\dots,y^{i_m}, 0}
	\end{align}
	and
	\begin{align}
		\label{phi1} \ket{\phi_1} = \frac{1}{\sqrt{\mu}} \sum_{i_{1} \cdots i_{m}} \Ket{y^{i_1},\dots,y^{i_m}, 1}
	\end{align}
	counting the cases in which $\delta^{i_{1} \cdots i_{m}} = 1$, we obtain:
	\begin{align}
		\label{mu} \mu &= \frac{N!}{m!(N-m)!},
		\\
		\label{nu} \nu &= N^{m} - \mu,
		\\
		\label{theta} \theta &= \arcsin\left( \sqrt{\frac{\mu}{N^{m}}} \right).
	\end{align}
	Notice that $\ket{\phi_1}$  is a superposition of all the sequences $\left\lbrace y^{i_1},\dots,y^{i_m} \right\rbrace$ such that the $y^{i_1} < \cdots < y^{i_m}$.
	
	\item Define $p \in \mathbb{N}$ and apply Grover's algorithm $p$ times (denoted by $\mathcal{G}(p)$) on the last qubit in order to amplify the amplitude of the state $\ket{1}$:
	\begin{align}
		\notag \label{mpfinalstate}	\ket{\psi_{3}} &= I \otimes \mathcal{G}(p) \ket{\psi_{2}}
		\\
		&= \cos\left[ \left(2p +1\right) \theta \right]  \ket{\phi_0} + \sin\left[ \left(2p +1\right) \theta \right] \ket{\phi_1},
	\end{align}

	\item Measure the first $n$ qubits; that is, the ones that correspond to the register $y^{i_{1}}$ (see figure \ref{fig:1}). This last step is omitted when using the $(m,p)$ algorithm as a subroutine.
\end{enumerate}

\begin{figure*}[t]
	\begin{centering}
		\includegraphics[width=0.47\linewidth]{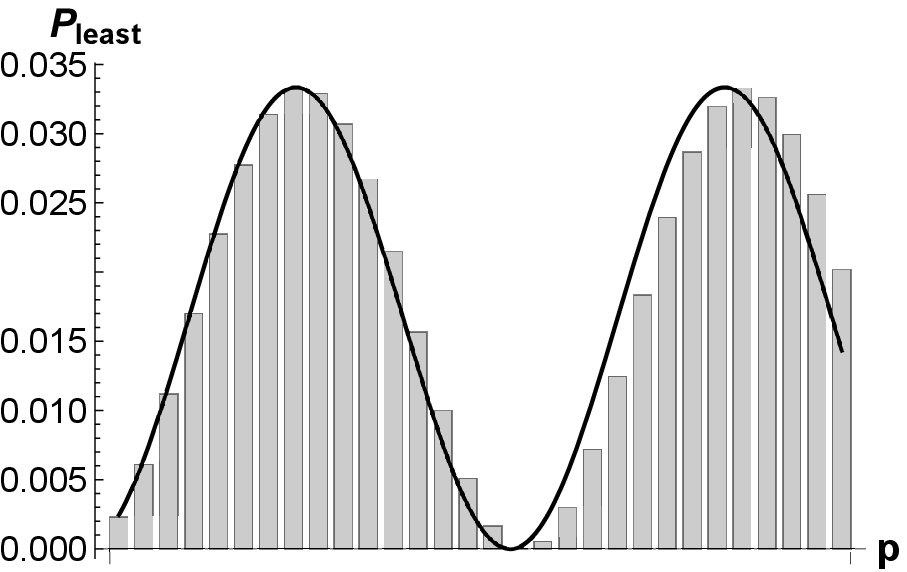} \qquad
		\includegraphics[width=0.47\linewidth]{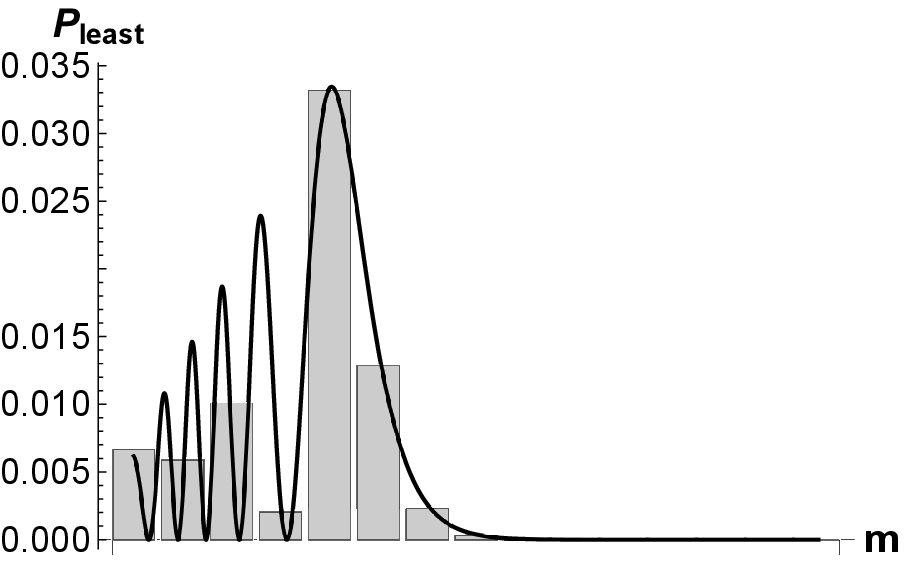}
		\par\end{centering}
	\caption{\label{fig:3} Left: Probability of measuring the least element in $Y$ as a function of $p$ using $m = 5$. Right: Probability of measuring the least element in $Y$ as a function of $m$ using $p = 8$. In both cases the exact value (eq. \eqref{pmax}) is represented by the bar chart and the approximated value (eq. \eqref{pmaxap}) by the continuous line. $N=150$ was used.}
\end{figure*}

At the end of the algorithm, if $p$ is chosen such that $\sin\left[ \left(2p +1\right) \theta \right] \approx 1$, then the probability distribution of the measurement result is given by
\begin{equation}
	\label{pd}	P(x) = \begin{cases}
		\frac{m(x-1)!(N-m)!}{(x-m)!N!} \qquad \text{if } m\leq x \leq N,
		\\
		0 \hspace{2.7cm} \text{if } 1\leq x < m,
	\end{cases}
\end{equation}
where $x$ represents the order label of the elements in $Y$, $x = N$ corresponding to the least element and $x = 1$ to the largest one. Figure \ref{fig:2} shows a bar chart of distribution \eqref{fig:2} for different values of $m$.

A feature of this sorting algorithm is that, by tuning its parameters, it can be adapted to specific quantum memory ($m$) and circuit depth ($p$) requirements; for example, if we wished to run it on a quantum computer with low memory but high circuit depth capacity, we could tune $m$ to be restricted by the hardware and left $p$ to depend on $m$. To analyze the relation between these two parameters, let us focus only on the second term in expression \eqref{mpfinalstate}. In this case, the probability of measuring the least element in $Y$ at the end of the algorithm is given by
\begin{equation}
	\label{pmax} P_{\text{least}} = \frac{m}{N} \cdot \sin^{2}\left[ (2p+1) \arcsin\left( \sqrt{\frac{N!}{m!(N-m)!N^{m}}} \right) \right].
\end{equation}
If we assume $N \gg m$ and $\arcsin\left( \frac{1}{\sqrt{m!}} \right) \approx \frac{1}{\sqrt{m!}}$, equation \eqref{pmax} simplifies to
\begin{equation}
	\label{pmaxap} P_{\text{least}} \approx \frac{m}{N} \cdot \sin^{2}\left[ (2p+1) \sqrt{\frac{1}{m!}} \right].
\end{equation}
Equation \eqref{pmaxap} can be used to tune the two parameters for specific requirements.

Figure \ref{fig:3} shows the relation between $m$ and $p$ by plotting expressions \eqref{pmax} and \eqref{pmaxap} as both a function of $p$ with a fixed $m$ and a function of $m$ with a fixed $p$. We see that, for a fixed $m$, we might choose $p$ to be the value corresponding to the first maximum of $P_{\text{least}}$, while for a fixed $p$, we might choose $m$ to be the value that maximizes $P_{\text{least}}$, this of course under the condition $(2p+1) \sqrt{\frac{1}{m!}} \approx \frac{\pi}{2}$.

Regarding the complexity of the algorithm, this lies in the parameter $p$; that is, it is of complexity $O(p)$ (without taking into account the initialization of the circuit). The condition $(2p+1) \sqrt{\frac{1}{m!}} \approx \frac{\pi}{2}$ implies
\begin{equation}
p_{\text{optimal}} \approx \frac{\pi}{4} \sqrt{m!} - \frac{1}{2},
\end{equation}
which is the optimal value for $p$. If we wish to keep the quantum advantage of Grover's algorithm, we must restrict $p$ to be $O(\sqrt{N})$, or equivalently, restrict $m!$ to be $O(N)$. However, it is worth mentioning that $m$ can be considered independent of $N$, as it is only bounded by it, and that relatively small values of $m$ suffice for a good performance, as we will show in section \ref{ResD}.

\section{Quantum version of the k-NN algorithm based on the (m,p) sorting algorithm} \label{MP_kNN}

In this section we propose a new quantum version of the k-NN algorithm based on the $(m,p)$ sorting algorithm. As in the classical k-NN algorithm, consider a pattern $x^{\omega}$ to be classified and a training set of binary $n$-dimensional patterns $E = \left \lbrace (x^{1}, c^{1}), \dots, (x^{N}, c^{N}) \right \rbrace $. Then, our proposed algorithm is as follows:

\begin{enumerate}
	
	\item Prepare the initial state
	\begin{equation}
		\ket{\psi_{1}} = \ket{E_{\omega}} \otimes \Ket{E_{x}}^{\otimes m-1} \otimes \ket{0},
	\end{equation}
	where
	\begin{align}
		\ket{E_{\omega}} &= \frac{1}{\sqrt{N}} \sum_{i=1}^{N} \Ket{c^{i},x^{\omega},x^{i}},
		\\
		\ket{E_{x}} &= \frac{1}{\sqrt{N}} \sum_{i=1}^{N} \Ket{x^{i}}.
	\end{align}

	\item Apply a metric operator $\mathcal{M}(x^{\omega}, x^{i})$ that computes any defined distance between $x^{\omega}$ and all the patterns $x^{i}$, and encode it in the qubits corresponding to the $m$ states $\ket{x^{i}}$ (see figure \ref{fig:4}). As in Schuld's algorithm, in this work we use the Hamming distance, whose metric operator is implemented using $CNOT(x^{\omega}_{j},x^{i}_{j})$ and $X(x^{i}_{j})$ gates. At the end of this step the state of the system is
	\begin{align}
		\notag \ket{\psi_{2}} =&  \frac{1}{\sqrt{N}} \sum_{i=1}^{N} 	\Ket{c^{i},x^{\omega},d^{\, i}}
		\\
		&\otimes \left(\frac{1}{\sqrt{N}} \sum_{i=1}^{N} \Ket{d^{\,i}}\right)^{\otimes m-1} \otimes \ket{0},
	\end{align}
	where $d^{\,i}$ is defined as in equation \eqref{distance}.
	
	\item Apply the $(m,p)$ sorting algorithm to the last $mn+1$ qubits (see figure \ref{fig:4}); that is, to the states $\Ket{d^{\,i}}^{\otimes m} \otimes \ket{0}$. The final state of the algorithm is then given by
	\begin{align}
		\notag	\ket{\psi_{3}} &=
		\\
		\notag &\frac{\cos\left[ \left(2p +1\right) \theta \right]}{\sqrt{\nu}} \sum_{i_{1} \cdots i_{m}} \Ket{c^{i_1}, x^{\omega}, d^{i_1},\dots,d^{i_m},0}
		\\
		\label{final_state_mp_knn} &+ \frac{\sin\left[ \left(2p +1\right) \theta \right]}{\sqrt{\mu}} \sum_{i_{1} \cdots i_{m}} \Ket{c^{i_1}, x^{\omega}, d^{i_1},\dots,d^{i_m}, 1},
	\end{align}
	where $\mu, \nu$ and $\theta$ are the same as in expressions \eqref{nu}, \eqref{mu} and \eqref{theta} respectively.
	
	It is worth mentioning that the $(m,p)$ sorting algorithm requires that the set to be sorted has a strict total order relation; this does not always happen, as it is common to have two or more patterns that are at the same distance from the one to be classified. However, this type of relation can be artificially implemented; for example, if the considered metric operator $\mathcal{M}(x^{\omega}, x^{i})$ is the one corresponding to the Hamming distance, and two patterns $x^{i_{1}}$ and $x^{i_{2}}$ are at the same distance from $x^{\omega}$, then $d^{\,i_{1}} $ and $d^{\,i_{2}}$ will be two $n$-bit registers with the same amount of components equal to $1$. To implement the artificial strict total order, it can be considered that $d^{\, i_{1}} > d^{\, i_{2}}$ or $d^{\, i_{2}} > d^{\, i_{1}}$ at random. The multiple ways in which the strict total order can be implemented may slightly affect the performance of the proposed algorithm, as any tiebreaker rule also does in the classical k-NN algorithm.
	
	\begin{figure}[t]
		\begin{centering}
			\includegraphics[width=0.99\linewidth]{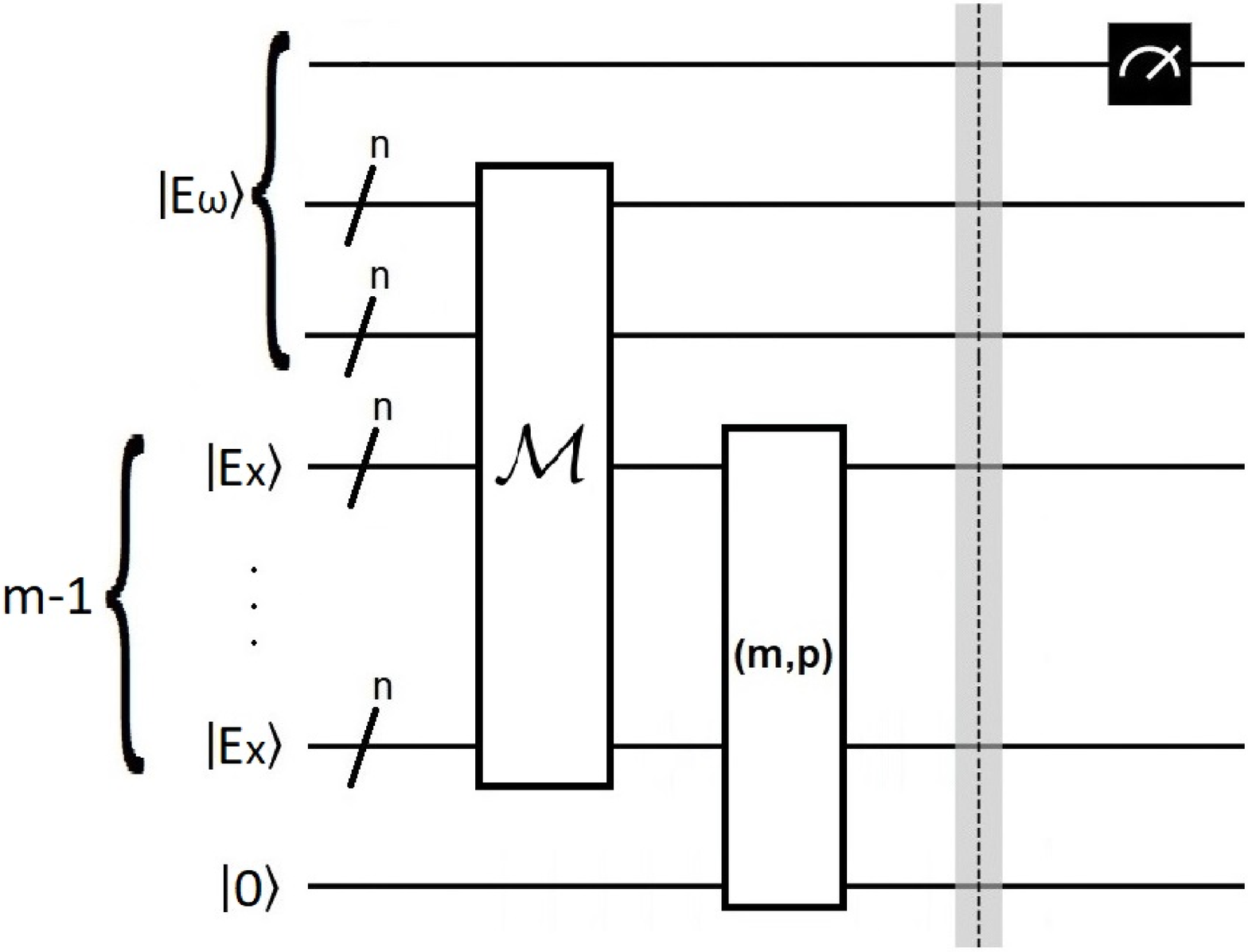}
			\par\end{centering}
		\caption{\label{fig:4} Quantum circuit of the quantum k-NN algorithm based on the $(m,p)$ sorting algorithm.}
	\end{figure}
	
	\item Measure the class qubits; that is, those corresponding to the state $\Ket {c^{i_{1}}} $ (see figure \ref{fig:4}), and store the obtained result.
	
	By appropriately choosing the parameters $m$ and $p$, the probability of measuring the class of the patterns nearest to $x^{\omega}$ is maximized. Being strict, the probability of measuring an arbitrary class $c$ at the end of the circuit is given by
	\begin{align}
		\notag P(c) =& \frac{\cos^{2}\left[ (2p+1) \theta \right]}{\nu} \sum_{x \in c} N_{0}(x)
		\\
		\label{prob_class_mp} &+ \frac{\sin^{2}\left[ (2p+1) \theta \right]}{\mu} \sum_{x \in c} N_{1}(x)
	\end{align}
	where the sum is performed on the order label $x$ of the distances corresponding to the patterns in class $c$, $N_{0}(x)$ is the number of times that the element with label $x$ appears in the position $d^{i_{1}}$ of the first sum in expression \eqref{final_state_mp_knn}, and $ N_{1}(x)$ is the number of times it appears in the second sum of the same expression:
	\begin{align}
		N_{0}(x) &= \begin{cases}
			N^{m-1} \hspace{3.35cm} \text{if } x < m, \\
			N^{m-1} - \frac{(x-1)!}{(m-1)!(x-m)!} \qquad \text{if } x \geq m.
		\end{cases}
		\\ \notag \\
		N_{1}(x) &= \begin{cases}
			0 \hspace{2.65cm} \text{if } x < m, \\
			\frac{(x-1)!}{(m-1)!(x-m)!} \qquad \text{if } x \geq m.
		\end{cases}
	\end{align}

	\item The algorithm is run $k$ times. The statistical mode of the $k$ obtained results is assigned to $x^{\omega}$.
\end{enumerate}

As in the quantum version of the k-NN algorithm proposed by Schuld et al., the value of $k$ is not restricted to the number of patterns in the training set, as it can be interpreted as the number of samples to be obtained in a random experiment with replacement, where the probabilities of measuring each class are given by the expression \eqref{prob_class_mp}.

Regarding the complexity of the algorithm, since the metric operator is $O(nm)$ and the $(m,p)$ sorting algorithm is $O(p)$, the overall complexity can be considered to be $O(knmp)$, once again, without taking the circuit initialization into account.

\section{Results and discussion} \label{ResD}

In this section we compare the performance of the classical k-NN, quantum version by Schuld et al. and the quantum $(m,p)$-based version. In order to do so, we use the famous Iris dataset \cite{Iris}, which consists of $150$ patterns and $4$ continuous numerical attributes separated in $3$ classes of $50$ patterns each: Iris Setosa, Iris Versicolor and Iris Virginica.

Since both quantum algorithms work with binary patterns, the attributes present in the Iris dataset must be binarized. For this, we decided to use the gray code \cite{gray}, whose main feature is that two consecutive numbers differ only in one digit, an ideal characteristic for comparing numerical attributes. The specific transformation went as follows: first, all the attributes were multiplied by $10$, resulting in all having integer values; then, each integer was transformed to its representation in the gray code; finally, all digits were separated into different attributes.

Quantum algorithms are inherently probabilistic, therefore, in order to avoid incorporating more sources of randomness in the comparison process, we decided to use the so-called \textit{leave-one-out} cross validation method \cite{LOO}, as one of its main features is determinism. Lastly, as the Iris dataset is perfectly balanced, for the comparison we decided to use accuracy (acc) as the performance measure:
\begin{equation}
	\text{acc} = \frac{\text{Number of correctly classified patterns}}{\text{Total number of patterns in the dataset}}.
\end{equation}

Initially, we used the values $k\in \left \lbrace 1,3,5,7,9,11,13 \right \rbrace$ for the three algorithms. In the case of the classical k-NN, the maximum accuracy was obtained for $k = 5$, having a value of $\text{acc} = 0.9533$. In the case of the quantum versions, given their probabilistic nature, the algorithms were run $50$ times for each value of $k$, thus obtaining an accuracy distribution. In the case of the quantum version proposed by Schuld et al., it was found that the average probability of measuring the last qubit as $0$ was approximately $2/3$ for the Iris dataset; thus, a value of $T=5k$ was fixed, resulting in a probability of not gathering $k$ class candidates to be less than $0.5\%$. In the case of the $(m,p)$-based quantum k-NN, the minimum optimum values were found to be $m=5$ and $p=8$.

\begin{figure}[t]
	\begin{centering}
		\includegraphics[width=1.00\linewidth]{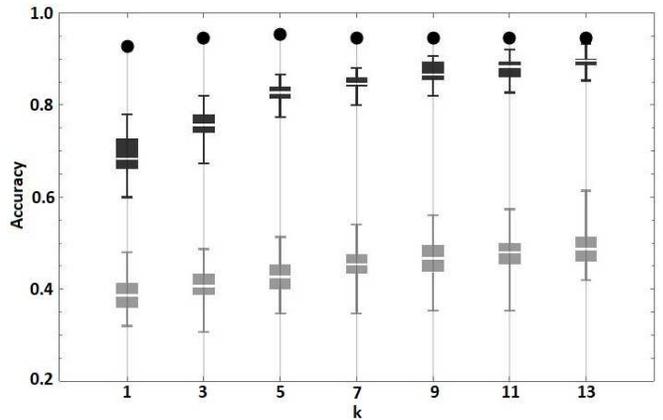}
		\par\end{centering}
	\caption{\label{fig:5} Accuracy obtained using the classical k-NN algorithm (black dots), the quantum version proposed by Schuld et al. (light gray box-whiskers) and the $(m,p)$-based quantum k-NN (dark gray box-whiskers) for $k \in \left\lbrace 1,3,5,7,9,11,13 \right\rbrace$. The box-whisker plots represent the accuracy distribution obtained from $50$ runs of the corresponding algorithm.}
\end{figure}

Figure \ref{fig:5} shows the performance results. The accuracy obtained with the classical algorithm is represented by a single dot, while the accuracy distribution obtained from the $50$ runs of each quantum algorithm is represented using a box and whisker plot. For all three algorithms, in the case of a tie among the $k$ class candidates, the smallest mode was selected. At first sight, it can be observed that the classical algorithm surpasses both quantum versions. However, it is also observed that, for every analyzed value of $k$, the accuracy distribution obtained with the quantum algorithm proposed by Schuld et al. is below the distribution obtained with the quantum k-NN based on the $(m,p)$ algorithm.

\begin{figure}[t]
	\begin{centering}
		\includegraphics[width=1.00\linewidth]{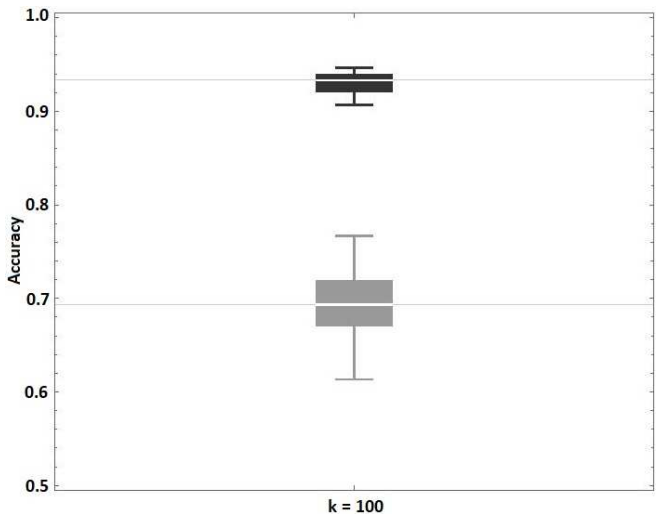}
		\par\end{centering}
	\caption{\label{fig:6} Accuracy distribution obtained using the quantum k-NN proposed by Schuld et al. (light gray box-whisker) and the $(m,p)$-based quantum k-NN (dark gray box-whisker) for $k=100$. The box-whisker plots represent the accuracy distribution obtained from $50$ runs of the corresponding algorithm.}
\end{figure}

On the other hand, as it was expected, the mean accuracy of both quantum algorithms increases with the value of $k$, which is a direct consequence of its interpretation as the number of samples in a random experiment with replacement. As an example of this, figure \ref{fig:6} shows the distributions obtained for $k = 100$. The maximum mean accuracy of both quantum algorithms can be calculated through the probabilities of measuring each class for each pattern: if the maximum probability turns out to be that of the class to which the pattern belongs, then in the limit $ k \rightarrow \infty$ ($ k \gg 1 $ in practice) that pattern will be correctly classified, otherwise it will be incorrectly classified. Then, performing the described analysis, it is obtained that the maximum mean accuracy for the quantum k-NN proposed by Schuld et al. is
\begin{equation} 
	\text{acc}_{\text{Schuld}}^{\text{máx}} = 0.9066,
\end{equation}
while the maximum mean accuracy for the quantum k-NN based on the $(m,p)$ algorithm is
\begin{equation} 
	\text{acc}_{(m,p)}^{\text{máx}} = 0.9466.
\end{equation}
In both cases, the quantum accuracy is below the maximum  obtained with the classical algorithm; nevertheless, in the classical case, $k$ is strongly tied to the number $N$ of patterns in the training set (it is recommended to search up to $k = O(\sqrt{N})$ \cite{RyD1}), whereas in quantum algorithms it is not. On the other hand, the class probability distribution of the $(m,p)$-based k-NN algorithm (equation \eqref{prob_class_mp}) seems more suitable for distance-based classification problems than the corresponding distribution of Schuld's quantum version (equation \eqref{prob_class_s}); thus, by using the $(m,p)$-based k-NN algorithm, it is possible to obtain a relatively high accuracy with few measurements, regardless of the number of patterns $N$.

\section{Conclusions} \label{C}

In this work, two quantum algorithms were proposed: the $(m,p)$ sorting algorithm, which sorts the elements of an array via their amplitude in a quantum superposition, and a new quantum version of the k-NN algorithm based on it. Furthermore, using the famous Iris dataset, the performance of our proposed $(m,p)$-based k-NN was compared with the performances of both the classical k-NN and the quantum version proposed by Schuld et al. \cite{Int13}.

Regarding performance, from the presented results it can be concluded that the proposal of this work within the area of quantum machine learning, that is, the quantum k-NN based on the $(m,p)$ algorithm, is superior to one of its main competitors, that is, the quantum k-NN proposed by Schuld et al. Furthermore, by tuning its parameters, the $(m,p)$-based k-NN can be adapted to quantum computers with limited memory or limited circuit depth.

Both the quantum version of the k-NN algorithm proposed by Schuld et al. and the quantum version proposed in this work, have the potential to be more efficient than the classical algorithm, provided that a more efficient way of initializing a quantum circuit is found. Regarding this issue, solutions such as receiving the initial state from a quantum memory or using efficient initialization protocols \cite{Sk-NN2} have been proposed.


\acknowledgments
L.F. Quezada acknowledges financial support from CONACYT México, from November 2020 to October 2021.

\end{document}